\documentclass[10pt]{article}

\usepackage{amsmath,amssymb,amsthm}
\usepackage{amsxtra}
\usepackage{amsfonts}
\usepackage{amssymb}
\usepackage{url}
\usepackage{hyperref}
\usepackage{esint}
\usepackage{eucal}
\usepackage{mathrsfs}
\usepackage{cite}
\usepackage{graphicx,graphics}
\usepackage{stackengine}
\setstackgap{S}{1pt}

\newcommand\labsect[1] {\label{sect:#1}}

\newcommand\eq[1] {(\ref{#1})}
\newcommand{\bfm}[1]{\mbox{\boldmath ${#1}$}}

\newcommand{\nonum}{\nonumber \\}
\newcommand{\beqa}{\begin{eqnarray}}
\newcommand{\eeqa}[1]{\label{#1}\end{eqnarray}}
\newcommand{\beq}{\begin{equation}}
\newcommand{\eeq}[1]{\label{#1}\end{equation}}
\newcommand{\Grad}{\nabla}
\newcommand{\Div}{\nabla \cdot}
\newcommand{\Curl}{\nabla \times}

\newcommand{\Imag}{\mathop{\rm Im}\nolimits}
\newcommand{\R}{\mathbb{R}}

\newcommand{\lang}{\langle}
\newcommand{\rang}{\rangle}
\newcommand{\Md}{\partial}

\newcommand{\Ga}{\alpha}

\newcommand{\Ge}{\epsilon}
\newcommand{\Gve}{\varepsilon}

\newcommand{\Gk}{\kappa}

\newcommand{\Gl}{\lambda}

\newcommand{\Gm}{\mu}

\newcommand{\Gt}{\theta}

\newcommand{\Gr}{\rho}
\newcommand{\Gvr}{\varrho}

\newcommand{\Go}{\omega}

\newcommand{\GO}{\Omega}

\newcommand{\BGa}{\bfm\alpha}

\newcommand{\BGe}{\bfm\epsilon}
\newcommand{\BGve}{\bfm\varepsilon}

\newcommand{\BGn}{\bfm\eta}
\newcommand{\BGm}{\bfm\mu}

\newcommand{\BGr}{\bfm\rho}

\newcommand{\BGs}{\bfm\sigma}

\newcommand{\BGy}{\bfm\psi}

\newcommand{\BGF}{\bfm\Phi}
\newcommand{\BGG}{\bfm\Gamma}

\newcommand{\BGT}{\bfm\Theta}

\newcommand{\CA}{{\cal A}}

\newcommand{\CE}{{\cal E}}

\newcommand{\CH}{{\cal H}}

\newcommand{\CJ}{{\cal J}}

\newcommand{\CP}{{\cal P}}
\newcommand{\CQ}{{\cal Q}}

\newcommand{\CT}{{\cal T}}
\newcommand{\CU}{{\cal U}}
\newcommand{\CV}{{\cal V}}

\newcommand{\BCL}{{\bfm{\cal L}}}

\newcommand{\BCP}{{\bfm{\cal P}}}
\newcommand{\BCQ}{{\bfm{\cal Q}}}

\newcommand{\BCY}{{\bfm{\cal Y}}}

\newcommand{\bpm}{\begin{pmatrix}}
\newcommand{\epm}{\end{pmatrix}}

\def\Ba{{\bf a}}
\def\Bb{{\bf b}}

\def\Bd{{\bf d}}
\def\Be{{\bf e}}

\def\Bh{{\bf h}}

\def\Bj{{\bf j}}
\def\Bk{{\bf k}}

\def\Bm{{\bf m}}
\def\Bn{{\bf n}}

\def\Bp{{\bf p}}
\def\Bq{{\bf q}}
\def\Br{{\bf r}}
\def\Bs{{\bf s}}
\def\Bt{{\bf t}}
\def\Bu{{\bf u}}
\def\Bv{{\bf v}}

\def\Bx{{\bf x}}

\def\BA{{\bf A}}

\def\BE{{\bf E}}

\def\BG{{\bf G}}

\def\BI{{\bf I}}
\def\BJ{{\bf J}}

\def\BL{{\bf L}}
\def\BM{{\bf M}}
\def\BN{{\bf N}} 

\def\BP{{\bf P}}
\def\BQ{{\bf Q}}
\def\BR{{\bf R}}
\def\BS{{\bf S}}

\def\BU{{\bf U}}
\def\BV{{\bf V}}
\def\BW{{\bf W}}

\def\BZ{{\bf Z}}

\usepackage{graphics}
\usepackage{amssymb}

\title{A unifying perspective on linear continuum equations prevalent in physics. Part VII: Boundary
  value and scattering problems}
\author{}
\date{}
\begin{document}
\maketitle
\vskip -.5cm
\centerline{\large Graeme W. Milton}
\centerline{Department of Mathematics, University of Utah, USA -- milton@math.utah.edu.}
\vskip 1.cm
\begin{abstract}
  We consider simply connected bodies or regions of finite extent in space or space-time and write conservation laws associated with the
  equations in Parts I-IV. We review earlier work where, for elliptic equations,
  the boundary value problem is reformulated as a problem in the abstract theory of composites and the associated effective operator is
  equated with the Dirichlet-to-Neumann map that governs the response of the body. The dielectric polarizability problem and acoustic and electromagnetic
  scattering by an inclusion are formulated as problems in the extended abstract theory of composites. The scattering response can be determined from appropriate intergrals over the inclusion. 
\end{abstract}
\section{Introduction}
\setcounter{equation}{0}
\labsect{i}
As in the previous parts \cite{Milton:2020:UPLI,Milton:2020:UPLII,Milton:2020:UPLIII,Milton:2020:UPLIV,Milton:2020:UPLV,Milton:2020:UPLVI}
we are interested in the plethora of linear physics equations expressible in
the form encountered in the extended abstract theory of composites:
\beq \BJ=\BL\BE-\Bs,\quad \BGG_1\BE=\BE,\quad\BGG_1\BJ=0,
\eeq{ad1}
where the field $\Bs$ is the source term, $\BGG_1$ is a projection
operator in Fourier space, while $\BL$ acts locally in space, or space-time, if the fields depend on time $t$.
The purpose of this part is threefold: to show that the form of $\BGG_1$ allows us to obtain a multitude of
conservation laws associated with the equations; following  Sections 2.4 and 2.5 of \cite{Grabovsky:2016:CMM}, and
Milton, in Chapter 3 of \cite{Milton:2016:ETC} to show that
boundary value problems for elliptic equations
can be reformulated in the form appropriate to the abstract theory of composites; and
extending \cite{Milton:2017:BCP} to show that not only acoustic scattering problems but also electromagnetic scattering problems can be
reformulated as problems in the abstract theory of composites.

Given any two fields $\BP_1(\Bx)$ and $\BP_2(\Bx)$ in this space of fields, we define the inner product of them
to be
\beq (\BP_1,\BP_2)=\int_{\mathbb{R}^3}(\BP_1(\Bx),\BP_2(\Bx))_{\CT}\,d\Bx,
\eeq{innp}
where $(\cdot,\cdot)_{\CT}$ is a suitable inner product on the space $\CT$
such that the projection $\BGG_1$ is selfadjoint with
respect to this inner product, and thus the space $\CE$ onto which
$\BGG_1$ projects is orthogonal to the space $\CJ$ onto which
$\BGG_2=\BI-\BGG_1$ projects. As in Part III, the integral should be over
$\mathbb{R}^4$ if the fields have a dependence on time $t=x_4$. 

We will not repeat the correspondence, given in Part I,
between the formulation given here and the
more standard  formulation, with the derivatives
of potentials explicitly entering the equations,
and with the solution involving resolvents.  

As in the preceding Parts, to avoid taking unnecessary transposes, and unless otherwise indicated,
we let $\Div$ or $\underline{\nabla}\cdot$ act on the first index of a field, and the action of $\Grad$ or $\underline{\nabla}$
produces a field, the first index of which is associated with $\Grad$ or $\underline{\nabla}$.

\section{Supercurrents and conserved quantities}
\setcounter{equation}{0}
\labsect{sc}
The fact that the differential constraints imply $\CE$ and $\CJ$ are orthogonal subspaces, means
that given a region $\GO$ some information must be communicated through the boundary $\Md\GO$
to ensure that the integral of $(\BJ,\BE)_{\CT}$ outside $\GO$ cancels the integral inside $\GO$.
If we consider electrical conductivity then we see that the integral of $\Bj\cdot\Be$ represents
the net power absorbed within $\GO$, and so there must a corresponding flow of power across
$\Md\GO$ matching the  net power generated outside $\GO$. Mathematically we can write
\beq \Bj\cdot\Be=\Div\BQ,\quad \text{with}\quad\BQ=-V\Bj,
\eeq{sc1}
where $V$ is the electrical potential associated with $\Be$: $\Be=-\Grad V$. Here $\BQ$ is a vector
valued flux, and
\beq W=\int_{\Md\GO}\BQ\cdot\Bn\,dS=\int_{\GO}\Div\BQ\,d\Bx=\int_{\GO}\Bj\cdot\Be\,d\Bx \eeq{sc2}
represents the flux of power through the boundary. Thus \eq{sc2} corresponds to conservation of
energy. Similar results of course apply to other conductivity type equations including
magnetotransport, the Faraday effect, and convective diffusion in an
incompressible stationary convective flow.
More generally, if we have $d\times n$ matrix valued fields $\BJ_0$ and $\BE_0$ such that
$\Div\BJ_0=0$ and $\BE_0=\Grad\Bu_0$, then we can write
\beq (\BE_0)^T\BJ_0=\Div\BQ,\quad \text{with}\quad\BQ=\Bu_0\otimes\BJ_0, \eeq{sc3}
where we call the $n\times d\times n$ component field $\BQ$ a supercurrent. Here $\Div$ contracts with the
second index of $\BQ$. The associated
conserved quantity
\beq \BW=\int_{\Md\GO}\BQ\cdot\Bn\,dS=\int_{\GO}\Div\BQ\,d\Bx=\int_{\GO}(\BE_0)^T\BJ_0\,d\Bx
\eeq{sc4}
has  $n\times n$ components. Here $\cdot\Bn$ contracts with the second index of $\BQ$. 
For example, for three dimensional elasticity, thermoelasticity and static poroelasticity, one has $n=3$
and $\BW$ is a second-order tensor with 9 independent components. For three dimensional piezoelectricity
$n=4$ so $\BW$ has 16 components. In this latter case there are four copies of $\Bk\otimes\Bk/k^2$ along
the block diagonal of $\BGG_1(\Bk)$ when the 3 components of the displacement field $\Bu(\Bx)$ are treated as
independent potentials. If one has addition couplings with magnetic fields then $n=5$ and $\BW$ has 25 components.
For quasistatic elasticity, with complex valued fields that one can separate into
real and imaginary parts, one has $n=6$, so $\BW$ has 36 real components.

There are more supercurrents and associated conserved quantities. In the two-dimensional case suppose $\Div\BJ_0=0$ in any
simply connected domain. Then we can write $\BJ_0=\BR_\perp\Grad\Bv_0$, where
\beq \BR_\perp=\bpm 0 & 1 \\ -1 & 0\epm \eeq{sc5}
is the matrix for a $90^\circ$ rotation. The supercurrent becomes the $2n\times 2 \times 2n$ component field,
\beq \BQ=\bpm \Bu_0 \\ \Bv_0 \epm \bpm \BR_\perp\Grad\Bu_0 & \BR_\perp\Grad\Bv_0 \epm
=\bpm \Bu_0\BR_\perp\Grad\Bu_0 & \Bu_0\BR_\perp\Grad\Bv_0 \\ \Bv_0\BR_\perp\Grad\Bu_0 & \Bv_0\BR_\perp\Grad\Bv_0\epm 
\eeq{sc5a}
and the conservation law reads as
\beqa \BW & = & \int_{\Md\GO}\BQ\cdot\Bn\,dS=\int_{\GO}\Div\BQ\,d\Bx
=\int_{\GO}\bpm (\BE_0)^T \\ -(\BR_\perp\BJ_0)^T \epm \bpm \BR_\perp\BE_0 & \BJ_0 \epm \,d\Bx \nonum
& = & \int_{\GO}\bpm (\BE_0)^T\BR_\perp\BE_0 & (\BE_0)^T\BJ_0 \\ -(\BJ_0)^T\BE_0 & (\BJ_0)^T\BR_\perp\BJ_0 \epm\,d\Bx.
\eeqa{sc6}
Thus the conserved quantity $\BW$ is a $2n\times 2n$ antisymmetric matrix. In a similar vein, in three-dimensions $\BA\BE_0=\BA\Grad\Bu_0$
has zero divergence for all antisymmetric matrices $\BA$ and accordingly the  $n\times n$ antisymmetric matrix
$(\BE_0)^T\BA\BE_0$ is a conserved quantity with $\Bu_0\BA\Grad\Bu_0$ as the corresponding supercurrent.

In general, the conservation laws follow from the structure of $\BGG_1(\Bk)$ and when $\BGG_1(\Bk)$ has a block
diagonal structure then there is at least one conserved quantity for each block. One may need to juggle the
constitutive law, interchanging some field components on the left and right sides of the constitutive law,
so that, for example, a block entry of $\BI-\Bk\otimes\Bk/k^2$ gets replaced by $\Bk\otimes\Bk$. If there are
$n$-appearances of the same block in $\BGG_1(\Bk)$ then one can expect at least $n^2$ conservation laws when
$\BL$ does not have any symmetries. 

The key identities giving rise to a conserved $n\times n$ matrix $\BW$ (where some comments on the associated values of $n$ will be
given shortly) include the following, where we assume a 3-dimensional space plus time if that is appropriate.
Associated with $\BZ(\Bk)$ is the key identity:
\beq  \bpm (\Grad \Bu)^T \\ \Bu \epm\cdot \bpm \BR \\ \Div\BR \epm=\Div\BQ,\quad\text{with}\quad
\BQ=\Bu\otimes\BR,
\eeq{sc8}
with an $n$-component vector field $\Bu$ and a $n\times d\times n$ component supercurrent $\BQ$ (where 
$\Div$ contracts with the second index of $\BQ$). It can be applied to:
the Oseen equations for incompressible fluid flow
around an object (n=3); steady viscous electron flow in graphene (n=3); perturbations of magnetohydrostatic equations (n=1);
 perturbations with Maxwell-Stresses present (n=4);
 time-harmonic acoustics (Helmholtz equation, n=1); elastodynamics (n=3);
 the Brinkman-Stokes-Darcy and compressible Oseen flow equations with
perturbations at constant frequency (n=3); 
 the Navier-Stokes incompressible fluid equations with constant
 frequency perturbations (n=3); the time harmonic linear thermoacoustic equations (n=4);
 the time harmonic multielectron Schr{\"od}inger equation
 (n=1); perturbations to the Nernst-Planck equations for the flow of charged particles in a fluid subject to an electric field,
 allowing for the electric field generated by the charged particles (n=1),
perturbations to diffusion of electrons and holes in semiconductors (n=1), the perturbed Boussinesq equations for a nonviscous 
fluid (n=3); and perturbed magnetohydrodynamic equations in an incompressible fluid (n=3).

The key identity associated with the time-harmonic Maxwell's equations is
\beq \bpm -\Curl\Br \\ \Br \epm \cdot\bpm \Bu \\ \Curl\Bu \epm=\Div\BQ\quad\text{with}\quad \BQ=\Bu\times\Br.
\eeq{sc9}

Associated with $\BG(\Bk,\Go)$ (or $\BS(\Bk,\Go)$, which is essentially the same) one has the key identity:
\beqa
\begin{pmatrix}
(\nabla \Bu)^T\\
i\partial \Bu/\partial t\\
\Bu
\end{pmatrix}\cdot
\begin{pmatrix}
\BR\\\Br\\ \nabla\cdot\BR +i\partial \Br/\partial t
\end{pmatrix} 
&=&(\nabla \Bu)^T\cdot\BR+i\frac{\partial \Bu}{\partial
t}\Br+\Bu\nabla\cdot\BR+i\Bu\frac{\partial \Br}{\partial t}\nonum
&=&\begin{pmatrix} \nabla \\ \partial/\partial t\end{pmatrix}\cdot\begin{pmatrix} \Bu\otimes\BR \\ i\Bu\otimes\Br
\end{pmatrix}.
\eeqa{sc10}
It can be applied to: heat, particle, or light diffusion
with or without convection (n=1); linearized reaction diffusion equations and predator-prey models with diffusion
and migration (n=2); Nernst-Planck equations for the flow of charged particles in a fluid subject to an electric field
(n=2); small perturbations to diffusion of electrons and holes in semiconductors (n=3); perturbed spintronic equations (n=3);
nuclear magnetic resonance with diffusion (n=3); perturbed Boussinesq equations for a nonviscous fluid (n=4);
full dynamic multielectron Schr{\"o}dinger equation (n=1); and Cosserat elasticity (n=6).

Associated with $\BN(\Bk,\Go)$ is the key identity
\beq \begin{pmatrix}
(\nabla \Bu)^T\\
-\partial \Bu/\partial t\epm\cdot \bpm \partial \Br/\partial t \\ \Div\Br \epm
=\begin{pmatrix} \nabla \\ \partial/\partial t\end{pmatrix}\cdot\BQ\quad\text{with}\quad
\BQ=\bpm \frac{\Md\Bu}{\Md t}\otimes\Br \\ -\Div\Bu\otimes\Br \epm.
\eeq{sc11}
It can be applied to: dynamic acoustics (n=1); elastodynamics (n=3); dynamic linear thermoelasticity (n=4);
dynamic piezoelectricity (n=4); dynamic poroelasticity (n=4); perturbations of two 
fluid models for superfluid liquids and gases (n=2); Cosserat elasticity (n=3); seepage in fissured rocks
(n=1); perturbed magnetohydrodynamic equations in an incompressible fluid (n=6).

For the dynamic Maxwell equation, the key identity is
\beq \begin{pmatrix}
-\Bh\\
\Bd
\end{pmatrix}\cdot\begin{pmatrix} \Bb\\ \Be\end{pmatrix}  = \begin{pmatrix} \nabla \\ \partial/\partial t\end{pmatrix}
\bpm \Bh\times\BGF- V \Bd \\ \Bd\cdot\BGF \epm,\quad
\eeq{sc12}
where we have used
\beq
\begin{pmatrix} \Bb\\ \Be\end{pmatrix}=\BGT\begin{pmatrix}\BGF \\ V \end{pmatrix}, \quad \BGT^{\dagger}\begin{pmatrix}
-\Bh\\
\Bd
\end{pmatrix}=0,\quad\text{with}\quad \BGT=\begin{pmatrix}\Curl & 0\\
                         -\frac{\Md}{\Md t} & -\Grad\end{pmatrix}.
\eeq{sc13}

Associated with $\BV(\Bk)$,  with $\Div\BR=\Div(\Div\BM)$, and $\BR$ and $\BM$ being $d\times n$ and $d\times d\times n$
index objects, is the key identity
\beq
\bpm \BR \\ \BM \epm\cdot \bpm \Grad\Bu \\ \Grad\Grad \Bu \epm
=\Div\BQ,\quad\text{with}\quad \BQ=(\Grad\Bu)^T\BM+\Bu\otimes(\BR-\Div\BM),
\eeq{sc15}
in which $\BQ$ has $n\times d\times n$ indices and the divergences in $\Div\BQ$ and $\Div\BM$ act on the second index.
It can be applied to: higher order gradient conductivity (n=1);
elasticity (n=3); flexoelectricity (n=4); flexomagnetic equations (n=4);
and flexomagnetoelectricity (n=5).

The elastodynamics of thin plates (n=1) has the key identity,
\beq \bpm \Md \BM/\Md t \\  -\Div(\Div\BM) \epm\cdot \bpm -\Grad\Grad v \\  \Md \BM/\Md t \epm
=\begin{pmatrix} \nabla \\ \partial/\partial t\end{pmatrix}\cdot\BQ\quad\text{with}\quad
\BQ=\bpm -\frac{\Md v}{\Md t}\Div\BM-\frac{\Md \BM}{\Md t}\Grad v \\
(\Div\BM)\cdot(\Div v) \epm
\eeq{sc16}
while the Mindlin equations for moderately thick plates has the key identity:
\beq \begin{pmatrix} \Md \BM/\Md t \\  \Md \Bt/\Md t \\ \Bt-\Div\BM \\ \Div\Bt \end{pmatrix}\cdot
\begin{pmatrix} \Grad\dot{\BGy} \\ \dot{\BGy} - \Grad \dot{w} \\ \dot{\BGy} \\ \Md \dot{w}/\Md t \epm
  = \begin{pmatrix} \nabla \\ \partial/\partial t\end{pmatrix}\cdot\BQ\quad\text{with}\quad
  \BQ=\bpm -\dot{w}\frac{\Md\Bt}{\Md t}-\BM\frac{\dot{\BGy}}{\Md t}\\
  \dot{w}\Div\Bt-\Bt\cdot\dot{\BGy}+\BM:\Grad\dot{\BGy} \epm.
\eeq{sc17}

Some comments are needed to explain the values of $n$ provided. The form of $\BL$ in the constitutive law
may force $\Div\BR=0$ in \eq{sc8}, in which case this identity reduces to the identity in \eq{sc4}. Similarly
if the constitutive law forces $\Br=0$ in \eq{sc10} then the identity reduces to that in \eq{sc8}.
The converse is
true too: an identity of the form \eq{sc4} can be treated as a special case of an identity of the form \eq{sc8}
and accordingly add to the value of $n$ associated with identities of the form \eq{sc8} in the equations.
Also an identity of the form \eq{sc8} can be treated as a special case of an identity of the form \eq{sc10}
and likewise add to the value of $n$ associated with identities of the form \eq{sc10} in the equations.
An illustrative example of both these points are the perturbed magnetohydrodynamic equations in an incompressible 
fluid, where one has
\beq  \bpm -\BGs'-\BGn(\Bj') \\ 0 \\ -\Div (\BGs'+\BGn(\Bj')) \\ 0 \\ \Gr'\BI \\ \Bp' \\ \Div\Bp'+\Md \Gr'/\Md t \epm\cdot
\bpm \Grad\Bb' \\ \Md \Bb'/\Md t \\ \Bb'\\ \Grad\Div\Bv' \\ \Grad\Bv' \\ \Md \Bv'/\Md t \\ \Bv'\epm=
 \bpm -\BGs'-\BGn(\Bj') \\ 0 \\ -\Div (\BGs'+\BGn(\Bj')) \\ \Gr'\BI \\ \Bp' \\ \Div\Bp'+\Md \Gr'/\Md t \epm\cdot
 \bpm \Grad\Bb' \\ \Md \Bb'/\Md t \\ \Bb'\\ \Grad\Bv' \\ \Md \Bv'/\Md t \\ \Bv'\epm
 =\begin{pmatrix} \nabla \\ \partial/\partial t\end{pmatrix}\cdot\BQ\quad\text{with}\quad
\BQ=\bpm \frac{\Md\Bu}{\Md t}\otimes\Br \\ -\Div\Bu\otimes\Br \epm.
\eeq{sc18}
where $\Bu$ has 6 components, being the 3 components of $\Bb'$ plus the 3 components of $\Bv'$,
while $\Br$ and $\BR$ have 6 and 18 components, respectively, being three $0$'s plus the three components of $\Bp'$,
and $\BGs'+\BGn(\Bj')$ plus those of $\Gr'\BI$ (giving totals of 3 and 7 independent nonzero
components, respectively). Thus we obtain an identity of the form \eq{sc10} with
$n=6$.

Three more points are to be emphasized: we only considered one solution to the equations and if there are $m$ solutions
(possibly with different tensors $\BL(\Bx)$) then
$n$ changes to $nm$; symmetries and zero entries of $\BL$ may reduce the number of independent components of $\BW$,
thus, for example, if $\BL$ is symmetric, then so too is $\BW$, and in time harmonic acoustics $\BR$ in \eq{sc8}
is of the form $-P\BI$, so $\BW$ only has 4 independent entries rather than 9; and the
entries of $\BQ$ might not be experimentally measurable, an example being the dynamic electromagnetic equations where
the potential $\BGF$ entering \eq{sc12} is not directly measurable.

\section{Boundary value problems for elliptic equations}
\setcounter{equation}{0}
\labsect{bv}
That boundary value problems could be formulated as problems in the abstract theory of composites, was
recognized independently by Grabovsky, in Sections 2.4 and 2.5 of \cite{Grabovsky:2016:CMM}, and
Milton, in Chapter 3 of \cite{Milton:2016:ETC}. The simplest example is that for electrical conductivity,
where, for a simply connected body $\GO$, one may take
\begin{itemize}
\item  $\CU'$ to consist of gradients of harmonic potentials $\Bu=-\Grad V$ with $\Grad^2 V=0$ in $\GO$,
\item  $\CE'$ to consist of fields $\Be =-\Grad V$ such that $\Be$ vanishes outside $\GO$ (implying $V$ is a constant on $\Md\GO$),
\item  $\CJ'$ to consist of fields $\Bj$ satisfying $\Div\Bj=0$ that vanish outside $\GO$ (implying $\Bn\cdot\Bj=0$ on $\Md\GO$).
\end{itemize}
The inner product between fields $\Bp_1(\Bx)$ and $\Bp_2(\Bx)$ is taken to be the standard $L^2$ inner product:
\beq (\Bp_1,\Bp_2)=\int_{\GO}\Bp_1(\Bx)\cdot\overline{\Bp_2(\Bx)}\,d\Bx.
\eeq{bvp1}
With respect to this inner product, a simple integration by parts shows that the three spaces $\CU'$, $\CE'$ and $\CJ'$ are
orthonormal. Note that fields $\Bu\in\CU'$ can be identified either by the boundary values of the associated potential $V$,
or equivalently by the tangential component of $\Bu=-\Grad V$, or by the normal component $\Bn\cdot\Bu$ of $\Bu$ that
represents a flux. The conductivity equations inside $\GO$ can now be written as 
\beq \Bj(\Bx)=\BGs(\Bx)\Be(\Bx), \quad \Bj\in\CU\oplus\CJ,\quad \Be\in\CU\oplus\CE,
\eeq{bvp2}
where $\BGs(\Bx)$ is the conductivity tensor. The standard Dirichlet boundary problem is then interpreted as a problem in the
abstract theory of composites: given $\Bu_{\Be}\in\CU$ (identified by the boundary values of the associated potential)
solve \eq{bvp2} subject to the constraint that $\BGG_0\Be=\Bu_{\Be}$,
where $\BGG_0$ is the projection onto $\CU$. Similarly, the standard Neumann boundary problem becomes: given $\Bu_{\Bj}\in\CU$
(identified by the boundary values of the associated flux $\Bn\cdot\Bu_{\Bj}$)
solve \eq{bvp2} subject to the constraint that $\BGG_0\Bj=\Bu_{\Bj}$. Note that solving these problems can be made without
reference to the boundary values of the potential and without reference to the boundary value of the flux $\Bn\cdot\Bu_{\Bj}$:
one just solves the problems with given fields $\Bu_{\Be}$ and $\Bu_{\Bj}$ that are gradients of harmonic potentials.
As $\BGG_0\Bj=\Bu_{\Bj}$ depends linearly on $\BGG_0\Be=\Bu_{\Be}$, one can write
\beq \Bu_{\Bj}=\BCL_*\Bu_{\Be} \eeq{bvp3}
where $\BCL_*:\CU'\to\CU'$ is the associated effective operator. As $\Bu_{\Be}$ can be identified with its associated potential,
and $\Bu_{\Bj}$ with its associated boundary flux, we see that $\BCL_*$ represents the Dirichlet-to-Neumann map characterizing the electrical
response of the body. If there are current sources within $\GO$, then \eq{bvp2} should be modified to
\beq  \Bj'(\Bx)=\BGs(\Bx)\Be(\Bx)-\Bs(\Bx), \quad \Bj\in\CU\oplus\CJ,\quad \Be\in\CU\oplus\CE,
\eeq{bvp4}
where $\Bs(\Bx)$ must be chosen so $\Div\Bs$ equals the current source, and $\Bj=\Bj'+\Bs$ is the electrical current.
In this formulation of boundary value problems for
electrical conductivity the projections $\BGG_0'$, $\BGG_1'$, and $\BGG_2'$ onto $\CU'$, $\CE'$, and $\CJ'$
no longer have a simple interpretation in
Fourier space. Nevertheless the action of $\BGG_1'$ can be computed using fast Poisson solvers
(Kaushik Bhattacharya, private communication): given $\Bp$, to obtain
$\BGG_1'\Bp$ one can use a fast Poisson solver to find $V$ satisfying
\beq \nabla^2 V=-\Div\Bp,\quad V=0\quad\text{on}\quad\Md\GO, \eeq{fps}
and then $\Be=-\Grad V=\BGG_1'\Bp$. Accordingly, rapidly converging series expansions can still be applied to solve \eq{bvp2}.

More generally, for other elliptic equations, we may follow Chapter 3 of \cite{Milton:2016:ETC} and take
\begin{itemize}
\item  $\CU'$ to consist of fields $\BU$ such that $\BL_0^{-1/2}\BU$ has a square-integrable
  extension outside $\GO$ giving a field in $\CE$, and
  $\BL_0^{1/2}\BU$ has a different  square-integrable extension outside $\GO$ giving a field in $\CJ$, 
\item  $\CE'$ consists of fields in $\BL_0^{1/2}\CE$ that are zero outside $\GO$,
\item  $\CJ'$ consists of fields in $\BL_0^{-1/2}\CJ$ that are zero outside $\GO$,
\end{itemize}
as our three orthogonal subspaces. Given any two fields $\BP_1(\Bx)$ and $\BP_2(\Bx)$ in this space of fields,
we define the inner product of them to be
\beq (\BP_1,\BP_2)=\int_{\GO}(\BP_1(\Bx),\BP_2(\Bx))_{\CT}\,d\Bx,
\eeq{bvp5}
where $(\cdot,\cdot)_{\CT}$ is a suitable inner product on the space $\CT$ such that $\CE'$ and $\CJ'$ are orthogonal
spaces.

This formulation also applies to time harmonic equations in lossy media after one first employs the Cherkaev-Gibiansky
transformation \cite{Cherkaev:1994:VPC} to obtain an equivalent problem where $\BL(\Bx)$ is Hermitian and positive definite. A natural choice of $\BL_0$
is produced by applying the  Cherkaev-Gibiansky transformation to a tensor corresponding to a
homogeneous lossy medium.
In particular, if the loss in the original problem is small, then the transformation should be applied to a tensor
with small loss.

Another approach is to reformulate the equations as a ``$Y$-problem''. ``$Y$-problems'' are reviewed in Chapters 19 and 20 of \cite{Milton:2002:TOC} and in
Chapter 2 of \cite{Milton:2016:ETC}. 
For simplicity, let us just consider
the conductivity problem. We define
\begin{itemize}
\item  $\CV''$ to consist of fields, going to zero as $|\Bx|\to\infty$, that are  gradients of harmonic potentials outside $\GO$,
  and which are zero inside $\GO$
\item  $\CH'' $  to consist of fields that are zero outside $\GO$
\item  $\CE''$ to consist of fields $\Be =-\Grad V$ in $\CV''\oplus\CH''$.
\item  $\CJ''$ to consist of fields $\Bj$ in $\CV''\oplus\CH''$ such that $\Div\Bj=0$.
  \end{itemize}
  Given $\Bv_e\in\CV''$ the ``$Y$-problem'' consists of finding fields $\Bv_j\in\CV''$ and $\Be, \Bj\in\CH''$ such that
  \beq \Bj(\Bx)=\BGs(\Bx)\Be(\Bx),\quad \Be, \Bj\in\CH'',\quad\text{and}\quad\Bv_j+\Bj\in\CJ'',\quad \Bv_e+\Be\in\CE''.
  \eeq{bvp20}
  Thus $\Bj$ and $\Be$ represent the current and electric field inside $\GO$, while $\Bv_j$ and $\Bv_e$ are current and electric
  fields outside $\GO$ that stimulate the fields inside $\GO$ and which measure the response of the body.
  Assuming the equations have a solution that is unique, for any given $\Bv_e$,
  the linear dependence of  $\Bv_j$ on $\Bv_e$, defines a ``$Y$-operator'',
  $\BCY_*:\CV''\to\CV''$, such that
  \beq \Bv_j=-\BCY_*\Bv_e. \eeq{bvp21}
  The fields in $\CV''$ (like the fields in $\CU'$) can be identified either by the boundary values that the associated potential
  $V(\Bx)$ takes on $\Md\GO$ or by the boundary values that the associated flux $\Bq=-\Bn\cdot\Grad V$ takes on $\Md\GO$. The
  minus sign in front of $\BCY_*$ is explained by the fact that
  \beq (\BCY_*\Bv_e,\Bv_e)=- (\Bv_j,\Bv_e)=-(\Bv_j+\Bj,\Bv_e+\Be)+(\Bj,\Be)=(\Bj,\Be)=(\BGs\Be,\Be).
  \eeq{bvp22}
  So the positive semidefiniteness of $\BGs$ implies that $\BCY_*$ defined in this way is positive semidefinite. The quantity on the
  right represents the power absorbed in $\GO$. Hence the quantity on the left of \eq{bvp21} has the interpretation of being
  the power supplied by the fields outside $\GO$. The Dirichlet-to-Neumann map characterizing the electrical
  response of the body is now represented by $\BCY_*$: by continuity of potential across $\Md\GO$ the boundary potential
  associated with $\Bv_e$ matches that associated with $\Be$, and by continuity of flux across $\Md\GO$, the boundary
  flux $\Bn\cdot\Bv_j$ matches the flux $\Bn\cdot\Bj$, where $\Bn$ is the outwards normal to $\Md\GO$.

\section{The dielectric polarizability problem}
\setcounter{equation}{0}
\labsect{asdpp}
Here we review the dielectric polarizability problem following Section 10.1 of \cite{Milton:2002:TOC} and Section 2
of \cite{Milton:2017:BCP}.
The constitutive law takes the form
\beq \underbrace{\Bd_0+\Bd^s(\Bx)}_{\Bd(\Bx)}=\BGve(\Bx)(\underbrace{\Be_0+\Be^s(\Bx)}_{\Be(\Bx)}), \eeq{die1}
where $\Bd_0$ and $\Be_0$ are constant fields, with $\Bd_0=\BGve_0\Be_0$,  while 
\beq \Be^s\in\CE,\quad \Bd^s\in\CJ, \eeq{die2}
in which $\CE$ is the space of square-integrable $d$-component vector fields that 
have zero curl, while $\CJ$ is the space of square-integrable $d$-component vector fields that 
have zero divergence. For simplicity the dielectric tensor sufficiently far away from the inclusion is assumed to be isotropic,
of the form $\BGe_0=\Ge_0\BI$, where $\Ge_0$ is a positive scalar. The constitutive law can be rewritten in the equivalent form:
\beq \Bd^s(\Bx)=\BGve(\Bx)\Be^s(\Bx)-\Bs(\Bx),\quad \text{where}\quad \Bs(\Bx)=\Bd_0-\BGve(\Bx)\Be_0=[\BGve_0-\BGve(\Bx)]\Be_0.
\eeq{die3}
Thus, with the constraints \eq{die2}, we arrive at a problem in the extended abstract theory of composites: given
$\Bs(\Bx)=[\BGve_0-\BGve(\Bx)]\Be_0$ find $\Bd^s$ and $\Be^s$ that solve \eq{die2} and \eq{die3}.

The electric potential $V(\Bx)$ outside any
sphere containing the inclusion has an expansion in spherical harmonics \cite{Jackson:1975:CEa}, the leading term of which is
\beq V^s(\Bx)=\Bb\cdot\Bx/(4\pi\Ge_0 r^3)  +\cdots{},
\eeq{die4}
and the associated electric field $\Be^s(\Bx)=-\Grad V^s(\Bx)$ is
\beq \Be^s=-\Grad V^s=-\Bb/(4\pi\Ge_0 r^3) +3\Bx(\Bb\cdot\Bx)/(4\pi\Ge_0  r^5)
 +\cdots{}.
\eeq{die5}
At large distances the dominant correction to the uniform field
comes from terms involving the vector $\Bb$; this vector is known as
the induced dipole moment. The factor of $4\pi\Ge_0$ is introduced so that
$\Bb$ has a direct interpretation as the first moment of the induced charge density. Since
the equations for the fields are linear, there must be
a linear relation between the induced dipole moment $\Bb$ and the applied
field $\Be_0$. This linear relation,
\beq \Bb=\BGa\Be_0, \eeq{die6}
defines the polarizability tensor $\BGa$ of the inclusion.

The vector $\Bb$ is determined by the integral
of the polarization field,
\beq \Bp(\Bx)=(\Ge(\Bx)-\Ge_0)\Be(\Bx)=\Bd(\Bx)-\Ge_0\Be(\Bx)=\Bd_0+\Bd^s(\Bx)-\Ge_0[\Be_0+\Be^s(\Bx)]
=\Bd^s(\Bx)-\Ge_0\Be^s(\Bx),
\eeq{die7}
over the volume of the inclusion. To see this we follow, for example, 
the argument given in Section 10.1 of \cite{Milton:2002:TOC}. Consider a ball $B_{r }$ of
very large radius $r$ containing the inclusion. Since the polarization
field is zero outside the inclusion, we can equate the integral of
the polarization field over the inclusion with the integral of
the polarization field over the ball $B_{r }$. Since the displacement field
$\Bd(\Bx)$ has zero divergence, and since $-\Be^s(\Bx)$ is the
gradient of the electrical potential $V^s(\Bx)$, it follows that
for any vector $\Bm$
\beqa \int_{B_{r }}\Bm\cdot\Bp(\Bx)\,d\Bx
&  = &\int_{B_{r }}\Bd^s(\Bx)\cdot\Grad(\Bm\cdot\Bx)\,d\Bx+\Ge_0\Bm\cdot\int_{B_{r }}\Grad V^s(\Bx)\,d\Bx
\nonum
& = &\int_{\partial B_{r }}(\Bm\cdot\Bx)\Bd^s(\Bx)\cdot\Bn
+\Ge_0 V^s(\Bx)\Bm\cdot\Bn \,dS
\nonum
& = &\Ge_0\Bm\cdot\int_{\partial B_{r }}V^s(\Bx)\Bn-\Bx(\Grad V^s(\Bx)\cdot\Bn)\,dS,
\eeqa{die8}
where $\Bn=\Bx/|\Bx|$ is the outward normal to the surface $\partial B_r$ of the
ball $ B_r$. When the radius $r$ of the ball $ B_r$ is sufficiently
large we can use the asymptotic formulas \eq{die4} and \eq{die5} to estimate
these integrals,
\beqa \int_{\partial B_{r }}\Bx(\Grad V^s(\Bx)\cdot\Bn)\,dS,
& \approx &-\int_{\partial B_{r }}2\Bx(\Bb\cdot\Bx)/(4\pi\Ge_0 r^4)\,dS
=\frac{2}{3\Ge_0}\Bb, \nonum
\int_{\partial B_{r }} V(\Bx)\Bn\,dS,
& \approx & \int_{\partial B_{r }}\Bx(\Bb\cdot\Bx)/(4\pi\Ge_0 r^4)\,dS
=\frac{1}{3\Ge_0}\Bb,
\eeqa{die9}
with these approximations becoming increasingly accurate as
the radius $r$ of the ball $B_{r }$ approaches infinity.
By subtracting these expressions and taking the limit as $r$ approaches
infinity we see that
\beq \int_{B_{r}}\Bp(\Bx)\,d\Bx=\Bb.
\eeq{die10}

The formula for quasistatic energy conservation takes the form:
\beq W=(\Go/2)\int_{\GO}(\Imag \BGve)\Be\cdot\overline{\Be}\,d\Bx= (\Go/2)\Imag\int_{\GO}\Bd\cdot\overline{\Be}\,d\Bx
=(\Go/2)\Imag\int_{\Md \GO}\overline{V}(\Bn\cdot\Grad V)\,dS,
\eeq{die11}
where $\GO$ is any simply connected region and $W$ represents the electrical power in $\GO$ converted into heat, averaged over time.
In particular, we can take $\GO$ as the ball $B_r$ of radius $r$ and use the asymptotic
expansion \eq{die4} to give
\beqa W & = &\lim_{r\to\infty}(\Go/2)\Imag\int_{\Md B_r}\Gve_0 \overline{V}\frac{\Md V}{\Md r}\,dS, \nonum
& = & \lim_{r\to\infty}(\Go\Gve_0/2)\Imag\int_{\Md B_r}
\left(\overline{\Be}_0\cdot\Bx-\frac{\overline{\Bb}_0\cdot\Bx}{4\pi\Ge_0 r^3}\right)
\left(\frac{\Be_0\cdot\Bx}{r}+\frac{\Bb_0\cdot\Bx}{2\pi\Ge_0 r^4}\right)\,dS, \nonum
& = &
\lim_{r\to\infty}(\Go/2)\Imag\int_{\Md B_r}\frac{2(\overline{\Be}_0\cdot\Bx)(\Bb_0\cdot\Bx)-(\Be_0\cdot\Bx)(\overline{\Bb}_0\cdot\Bx)}
{4\pi r^4}\,dS \nonum
  & =& (\Go/2)\Imag(\Bb_0\cdot\overline{\Be}_0) \nonum
  & = & (\Go/2)\Be_0\cdot(\Imag\BGa)\overline{\Be}_0,
  \eeqa{die12}
  where we have used the fact that $(\overline{\Be}_0\cdot\Bx)(\Be_0\cdot\Bx)$ is real and that integral of $\Bx\otimes\Bx$
  over the sphere surface $\Md B_r$ is $4\pi r^2\BI/3$. Thus, the imaginary part of $\BGa$ dictates the total amount of electrical power
  absorbed by the inclusion.

\section{Acoustic Scattering and Radiation}
\setcounter{equation}{0}
\labsect{as}
Here we follow the treatment in \cite{Milton:2017:BCP} for acoustic scattering as a prelude to
studying electromagnetic scattering.
\subsection{Framework}
\labsect{fr}

Let $P^a(\Bx)$ and $\Bv^a(\Bx)$ be the plane wave pressure and velocity fields that solve the acoustic equations in a homogeneous medium
with density $\Gr_0$ and bulk modulus $\Gk_0$, i.e.
\beq
\underbrace{\begin{pmatrix}-i\Bv^a \\ -i\Div\Bv^a \end{pmatrix}}_{\BJ^a}=
\underbrace{\begin{pmatrix}-(\Go\Gr_0)^{-1}\BI & 0 \\ 0 & \Go/\Gk_0\end{pmatrix}}_{\BL_0}
\underbrace{\begin{pmatrix}\Grad P^a \\ P^a \end{pmatrix}}_{\BE^a}.
\eeq{3.1}
Specifically, if $P^a(\Bx)=p^ae^{i\Bk_0\cdot\Bx}$ then these have the solution
\beq \BE^a=\begin{pmatrix}\Grad P^a \\ P^a \end{pmatrix}=\begin{pmatrix} i\Bk_0 p^a \\ p^a \end{pmatrix}e^{i\Bk_0\cdot\Bx}, \quad \quad
\BJ^a=\begin{pmatrix}-i\Bv^a \\ -i\Div\Bv^a \end{pmatrix}=\begin{pmatrix} -ip^a\Bk_0/(\Go\Gr_0) \\ p^a\Go/\Gk_0 \end{pmatrix}e^{i\Bk_0\cdot\Bx},
\eeq{3.2}
implying that $v^a=p^a\Bk_0/(\Go\Gr_0)e^{i\Bk_0\cdot\Bx}$ and that $\Bk_0$ must have magnitude $k_0=|\Bk_0|$ given by
\beq k_0=\sqrt{\Go^2\Gr_0/\Gk_0}. \eeq{3.3}

Given fields 
\beq \CP(\Bx)=\begin{pmatrix} \Bp(\Bx) \\ p(\Bx) \end{pmatrix},\quad \CP'(\Bx)=\begin{pmatrix} \Bp'(\Bx) \\ p'(\Bx) \end{pmatrix},\eeq{3.5a}
where $\Bp(\Bx)$ and $\Bp'(\Bx)$ are $d$-dimensional vector fields, and $p(\Bx)$ and $p'(\Bx)$ are scalar fields, 
we define the inner product
\beq (\CP',\CP)=\lim_{r_0\to\infty}\int_{t=0}^\infty w(t)(\CP',\CP)_{r_0t}\,\,dt, \eeq{3.5aa}
in which $w(t)$ is some smooth nonnegative weighting function, with say the properties that
\beq w(t)=0~~{\rm when}~~ t\leq 1/2~~{\rm or}~~t\geq 2,~~{\rm and}~~1=\int_{1/2}^2w(t)dt, \eeq{3.5ab}
and
\beq
(\CP',\CP)_r=\int_{B_r}\overline{\CP'(\Bx)}\cdot\CP(\Bx),\quad{\rm where}~
\overline{\CP'(\Bx)}\cdot\CP(\Bx)\equiv\overline{\Bp'(\Bx)}\cdot\Bp(\Bx)+\overline{p'(\Bx)}p(\Bx),\,\eeq{3.5b}
in which $B_r$ is the ball of radius $r$, with $\Md B_r$ being its spherical surface of radius $r$,  
and $\overline{a}$ denotes the complex conjugate of $a$ for any quantity $a$. We define $\CH^0$ as the space of 
fields $\CP^0$ such that the norm $|h\CP^0|=(h\CP^0,h\CP^0)^{1/2}$, with inner product given by \eq{3.5aa}, is finite for all scalar functions $h(\Bx)\in C_0^\infty(\R^d)$
(where $C_0^\infty(\R^d)$ is the set of all infinitely differentiable functions with compact support) and which additionally have the asymptotic behavior
\beq \CP^0(\Bx)=\frac{e^{ik_0|\Bx|}}{|\Bx|}\left\{\begin{pmatrix} \widehat{\Bx}R^s_\infty({\widehat{\Bx}}) \\ S^s_\infty({\widehat{\Bx}}) \end{pmatrix}+\mathcal{O}\left(\frac{1}{|\Bx|}\right)\right\}
+\frac{e^{-ik_0|\Bx|}}{|\Bx|}\left\{\begin{pmatrix} \widehat{\Bx}R^i_\infty({\widehat{\Bx}}) \\ S^i_\infty({\widehat{\Bx}})  \end{pmatrix}+\mathcal{O}\left(\frac{1}{|\Bx|}\right)\right\},
\eeq{3.baa}
for some complex scalar functions $R^s_\infty(\Bn)$, $S^s_\infty({\widehat{\Bx}})$, $R^i_\infty(\Bn)$  and $S^i_\infty(\Bn)$ defined on the unit sphere $|\Bn|=1$,  where $\widehat{\Bx}=\Bx/|\Bx|$. Here
the superscript $s$ is used because these field components will later be associated with the scattered field. The superscript $i$ is used because these field components will later be associated with 
incoming fields, though not the incoming fields associated with the incident fields $P^a$ and $\Bv^a$ as these will be treated separately. The subspace $\CH^0$
defined in this way has the property that if $\CP$ is in $\CH^0$ then so its complex conjugate $\overline{\CP}$.
We define $\CH^s$ as the space of fields $\CP^0\in\CH^0$ satisfying the 
condition that $R^i_\infty(\Bn)=S^i_\infty(\Bn)=0$ for all $\Bn$. Note that the norm $|\CP^0|=(\CP^0,\CP^0)^{1/2}$ is not finite for fields in $\CH^0$ 
if $R^s_\infty(\Bn)$, $S^s_\infty({\widehat{\Bx}})$, $R^i_\infty(\Bn)$  or $S^i_\infty(\Bn)$ is nonzero.

We are interested in solving
\beq
\underbrace{\begin{pmatrix}-i(\Bv^a+\Bv^s) \\ -i\Div(\Bv^a+\Bv^s) \end{pmatrix}}_{\BJ^a+\BJ^s}=
\underbrace{\begin{pmatrix}-(\Go\BGr)^{-1} & 0 \\ 0 & \Go/\Gk\end{pmatrix}}_{\BL(\Bx)}
\underbrace{\begin{pmatrix}\Grad (P^a+P^s) \\ (P^a+P^s) \end{pmatrix}}_{\BE^a+\BE^s},
\eeq{3.4}
where $P^s(\Bx)$ is the scattered pressure, $\Bv^s(\Bx)$ the associated scattered velocity, and $\BJ^s,\BE^s\in\CH^s$. Here the density
matrix $\BGr$ takes the value $\Gr_0\BI_d$ outside the inclusion and the value $\BGr_1(\Bx)$ inside the inclusion, while the bulk
modulus scalar $\Gk$ takes the value $\Gk_0$ outside the inclusion and the value $\Gk_1(\Bx)$ inside the inclusion. Due to viscoelasticity
(energy loss under oscillatory compression) it is quite natural to have a bulk modulus that is complex with a negative imaginary part.
We also allow for the density $\BGr_1(\Bx)$ to depend on the frequency $\Go$ and be anisotropic and possibly complex valued with a positive 
imaginary part, even with a negative real part, since this can be the case in metamaterials 
\cite{Schoenberg:1983:PPS,Auriault:1985:DCE,Auriault:1994:AHM,Zhikov:2000:EAT,Sheng:2003:LRS,Movchan:2004:SRR,Liu:2005:AMP,Milton:2006:CEM,Milton:2007:MNS,Torrent:2008:AMD, Smyshlyaev:2009:PLE,Buckmann:2015:EMU}. Alternatively, one can consider 
electromagnetic scattering off a cylindrical shaped inclusion (not necessarily with a circular cross-section) and then the transverse
electric and transverse magnetic equations are directly analogous to the two-dimensional acoustic equations. In that context it is
well known that both the electric permittivity tensor and magnetic permeability tensor can be anisotropic and complex valued, with positive
semidefinite imaginary parts.
 
Now using the relation \eq{3.1}, that $\BJ^a=\BL_0\BE^a$, we rewrite \eq{3.4} as
\beq \BJ^s(\Bx)=\BL(\Bx)\BE^s(\Bx)-\Bs(\Bx),\quad \Bs(\Bx)=(\BL_0-\BL(\Bx))\BE^a. \eeq{3.5}
We define $\CE^0$ as the space of all fields $\BE^0$ in $\CH^0$ of the form
\beq \BE^0=\begin{pmatrix}\Grad P^0(\Bx) \\ P^0(\Bx) \end{pmatrix}, \eeq{3.5ba}
for some scalar field $P^0(\Bx)$, and we define $\CJ^0$  as the space of all fields $\BJ^0$
in $\CH^0$ of the form
\beq\BJ^0= \begin{pmatrix}-i\Bv^0 \\ -i\Div\Bv^0 \end{pmatrix}, \eeq{3.5bb}
for some vector field $\Bv^0(\Bx)$. It looks like \eq{3.5}, together with the constraints that $\BJ_s\in\CJ^0$ and $\BE_s\in\CE^0$,
is a problem in the extended abstract theory of composites. This is not yet the case, as we will see shortly that $\CJ^0$ and $\CE^0$ are
not orthogonal spaces, and then we will remedy this defect. The nonorthogonality of $\CJ^0$ and $\CE^0$ is tied to the fact that radiation, and
hence energy, can escape to infinity. 
The fields $\BE^0$ and $\BJ^0$, being in $\CH^0$, have the asymptotic 
forms
\beqa \BE^0(\Bx)& = &\frac{e^{ik_0|\Bx|}}{|\Bx|}\left\{P^s_\infty({\widehat{\Bx}})\begin{pmatrix} ik_0\widehat{\Bx} \\ 1 \end{pmatrix}+\mathcal{O}\left(\frac{1}{|\Bx|}\right)\right\}
+\frac{e^{-ik_0|\Bx|}}{|\Bx|}\left\{P^i_\infty({\widehat{\Bx}})\begin{pmatrix} -ik_0\widehat{\Bx} \\ 1 \end{pmatrix}+\mathcal{O}\left(\frac{1}{|\Bx|}\right)\right\},\nonum
\BJ^0(\Bx) & = & \frac{e^{ik_0|\Bx|}}{|\Bx|}\left\{V^s_\infty({\widehat{\Bx}})\begin{pmatrix} -i\widehat{\Bx}/k_0 \\ 1 \end{pmatrix}+\mathcal{O}\left(\frac{1}{|\Bx|}\right)\right\}
+\frac{e^{-ik_0|\Bx|}}{|\Bx|}\left\{V^i_\infty({\widehat{\Bx}})\begin{pmatrix} i\widehat{\Bx}/k_0 \\ 1 \end{pmatrix}+\mathcal{O}\left(\frac{1}{|\Bx|}\right)\right\},
\eeqa{3.5bc}
implying, through \eq{3.5ba} and \eq{3.5bb}, that at large $|\Bx|$,
\beqa P^0(\Bx)& \approx & \frac{e^{ik_0|\Bx|}}{|\Bx|}P^s_\infty(\widehat{\Bx})+\frac{e^{-ik_0|\Bx|}}{|\Bx|}P^i_\infty(\widehat{\Bx}), \nonum
\Bv^0(\Bx)&\approx&\widehat{\Bx}\frac{e^{ik_0|\Bx|}}{k_0|\Bx|}V^s_\infty(\widehat{\Bx})-\widehat{\Bx}\frac{e^{-ik_0|\Bx|}}{k_0|\Bx|}V^i_\infty(\widehat{\Bx}),
\eeqa{3.5c}
in which the asymptotic field components satisfy:
\beq V^s_\infty=\frac{\Go}{\Gk_0}P^s_\infty=\frac{k_0}{\Go\Gr_0}P^s_\infty,\quad
V^i_\infty=\frac{\Go}{\Gk_0}P^i_\infty=\frac{k_0}{\Go\Gr_0}P^i_\infty.
\eeq{3.5ca}
The Sommerfeld radiation condition
\beq |\Bx|^{\tfrac{d-1}{2}}\left(\widehat{\Bx}\cdot\Grad P^0-ik_0 P^0\right)\to 0\quad\text{as}\quad |\Bx|\to 0,
  \eeq{som}
where $d=2,3$ denotes the dimensionality,
in fact implies that the $P^i_\infty(\widehat{\Bx})$ and $V^i_\infty(\widehat{\Bx})$ associated with
the actual scattered pressure and velocity are zero, but we keep these terms 
as we want to impose a ``constitutive law at infinity'' that forces $P^i_\infty(\widehat{\Bx})$ and $V^i_\infty(\widehat{\Bx})$ to be zero and thus
replaces the Sommerfeld radiation condition. Also we want to define the spaces $\CE^0$ and $\CJ^0$ so that if
$\BE^0\in\CE^0$ and $\BJ^0\in\CJ^0$ then so are their real and imaginary parts. 
We extend the definition of $P^s_\infty(\widehat{\Bx})$, and $V^s_\infty(\widehat{\Bx})$
to all of $\R^3$, excluding the origin, in the natural way by letting
\beq P^s_\infty(\Bx)=P^s_\infty(\Bx/|\Bx|),\quad V^s_\infty(\Bx)=V^s_\infty(\Bx/|\Bx|)\quad\text{for all  }\Bx\ne 0. \eeq{3.5d}
Then using the fact that $|\Bx|=\sqrt{\Bx\cdot\Bx}$ and $\widehat{\Bx}=\Bx/\sqrt{\Bx\cdot\Bx}$ we obtain
\beqa \Grad P^s(\Bx) & = & \frac{ik_0\Bx e^{ik_0|\Bx|}}{|\Bx|^2}P^s_\infty(\widehat{\Bx})
-\frac{\Bx e^{ik_0|\Bx|}}{|\Bx|^{3}}P^s_\infty(\widehat{\Bx})
+\frac{e^{ik_0|\Bx|}}{|\Bx|^2}\Bp^s(\Bx/|\Bx|), \nonum
\Div\Bv^s(\Bx) & = & \frac{ie^{ik_0|\Bx|}}{|\Bx|}V^s_\infty(\widehat{\Bx})
+(d-2)\frac{e^{ik_0|\Bx|}}{k_0|\Bx|^2}V^s_\infty(\widehat{\Bx})
+\frac{e^{ik_0|\Bx|}}{k_0|\Bx|^2}v^s(\Bx/|\Bx|),
\eeqa{3.5e}
where
\beq \Bp^s(\Bx/|\Bx|)=|\Bx|\Grad P^s_\infty(\Bx),\quad v^s(\Bx/|\Bx|)=\Bx\cdot\Grad V^s_\infty(\Bx)  \eeq{3.5f}
only depend on $\Bx/|\Bx|$, since $\Grad P^s_\infty(\Gl\Bx)=(1/\Gl)\Grad P^s_\infty(\Gl\Bx)$ 
and $\Grad V^s_\infty(\Gl\Bx)=(1/\Gl)\Grad V^s_\infty(\Gl\Bx)$ for all real $\Gl>0$. 
The dominant terms in the expressions in \eq{3.5e} are the first terms, which justifies those terms in \eq{3.5bc}
that involve $P^s_\infty({\widehat{\Bx}})$ and $V^s_\infty({\widehat{\Bx}})$. The terms that involve $P^i_\infty({\widehat{\Bx}})$ and $V^i_\infty({\widehat{\Bx}})$
are justified in a similar way by extending those functions to all of $\R^3$ except the origin.
Using integration by parts we have the key identity that
\beq (\BJ^0,\BE^0)_r=\int_{B_{r}}\BJ^0\cdot\overline{\BE^0}\,d\Bx = \int_{\Md B_{r}}-i\overline{P^0(\Bx)}\Bn\cdot\Bv^0(\Bx)\,dS.
\eeq{3.5faa}
From \eq{3.5c} we see that when $|\Bx|$ is large,
\beq \overline{P^0(\Bx)}\Bn\cdot\Bv^0(\Bx)\approx 
\frac{1}{k_0|\Bx|^2}\left[\overline{P^s_\infty({\widehat{\Bx}})}V^s_\infty({\widehat{\Bx}})-\overline{P^i_\infty({\widehat{\Bx}})}V^i_\infty({\widehat{\Bx}})
+e^{2ik_0r}\overline{P^i_\infty({\widehat{\Bx}})}V^s_\infty({\widehat{\Bx}})-e^{-2ik_0r}\overline{P^s_\infty({\widehat{\Bx}})}V^i_\infty({\widehat{\Bx}})\right].
\eeq{3.5fab}
The last two cross terms that involve $e^{2ik_0r}$ and $e^{-2ik_0r}$ obviously oscillate very rapidly with $r$ and will average to zero
in the integral \eq{3.5aa} involving the smooth weight function $w(t)$. Thus we get
\beqa (\BJ^0,\BE^0)& = & \lim_{r_0\to\infty}\int_{t=0}^\infty dt \frac{-iw(t)}{k_0r_0^2}\int_{B_{r_0t}}
\left[\overline{P^s_\infty({\widehat{\Bx}})}V^s_\infty({\widehat{\Bx}})-\overline{P^i_\infty({\widehat{\Bx}})}V^i_\infty({\widehat{\Bx}})\right]\,dS \nonum
& = &\frac{-i}{k_0}\int_{\Md B_{1}}
\left[\overline{P^s_\infty(\Bn)}V^s_\infty(\Bn)-\overline{P^i_\infty(\Bn)}V^i_\infty(\Bn)\right]\,dS.
\eeqa{3.5fa}
This lack of orthogonality of the subspaces $\CE^0$ and $\CJ^0$ can be remedied by introducing an auxiliary space $\CA$ of 
two-component vector fields $\Bq(\Bn)=[q_1(\Bn),q_2(\Bn)]$ defined, and square integrable, on the unit sphere $|\Bn|=1$. In a sense
$\CA$ represents fields ``at infinity''.
We then consider the Hilbert space $\CH$ composed of fields $[\CP,q_1,q_2]$, where $\CP\in\CH^0$ and $\Bq(\Bn)=[q_1(\Bn),q_2(\Bn)]\in\CA$.
In general, the field components $q_1(\Bn)$ and $q_2(\Bn)$ need not be related to the functions $R^s_\infty(\Bn)$, $S^s_\infty({\widehat{\Bx}})$, $R^i_\infty(\Bn)$  and $S^i_\infty(\Bn)$
appearing in the asymptotic expansion \eq{3.baa}. The inner product between two fields $\CQ=[\CP,q_1,q_2]$ and $\CQ'=[\CP',q_1',q_2']$ in $\CH$ is defined as
\beq \lang\CQ',\CQ\rang =(\CP',\CP)+\frac{1}{2k_0}\int_{|\Bn|=1}\overline{q'_1(\Bn)}q_1(\Bn)+\overline{q'_2(\Bn)}q_2(\Bn)\,dS. \eeq{3.5fb}
We define $\CE$ to consist of fields $\BE=[\BE^0,-iP^s_\infty+iP^i_\infty, P^s_\infty+P^i_\infty]\in\CH$, where $\BE^0\in\CE^0$ while $P^s_\infty(\Bn)$ 
and $P^i_\infty(\Bn)$ are those functions that enters its asymptotic form \eq{3.5bc}. 
We define $\CJ$ to consist of fields $\BJ=[\BJ^0, V^s_\infty+V^i_\infty, iV^s_\infty-iV^i_\infty]\in\CH$, where $\BJ^0\in\CJ^0$ while $V^s_\infty(\Bn)$ 
and $V^i_\infty(\Bn)$ are those functions that enters its asymptotic form \eq{3.5bc}. In each case the accompanying auxiliary fields are respectively
\beqa \Bq_{\BE}(\Bn)& = & \bpm -iP^s_\infty(\Bn)+iP^i_\infty(\Bn)\\ P^s_\infty(\Bn)+P^i_\infty(\Bn)\epm,~~{\rm and}\nonum
\Bq_{\BJ}(\Bn)& = & \bpm V^s_\infty(\Bn)+V^i_\infty(\Bn)\\ iV^s_\infty(\Bn)-iV^i_\infty(\Bn)\epm.
\eeqa{3.5fba}
The auxiliary fields have been defined in this way, in part, to ensure that if $\BE\in\CE$ and $\BJ\in\CJ$ then so too
are their complex conjugates, i.e., $\overline{\BE}\in\CE$ and $\overline{\BJ}\in\CJ$. 

Now the inner product of $\BE$ and $\BJ$ is 
\beq \lang \BJ,\BE\rang=(\BE^0,\BJ^0)+\frac{1}{k_0}\int_{\Md B_{1}}i\overline{P^s_\infty(\Bn)}V^s_\infty(\Bn)\,dS
+\frac{1}{k_0}\int_{\Md B_{1}}-i\overline{P^i_\infty(\Bn)}V^i_\infty(\Bn)\,dS
=0, 
\eeq{3.5fc}
which implies the orthogonality of the spaces $\CE$ and $\CJ$.

Of course since $\BE^s$ and $\BJ^s$ lie in $\CH^s$, rather than just $\CH^0$, the asymptotic components $P^i_\infty(\Bn)$ and $V^i_\infty(\Bn)$
are zero.
However let us remove this restriction and allow nonzero values of $P^i_\infty(\Bn)$ and $V^i_\infty(\Bn)$, that we will then show must be zero.
The associated fields $\BE=[\BE^s,-iP^s_\infty+iP^i_\infty, P^s_\infty+P^i_\infty]\in\CH$ and 
$\BJ=[\BJ^s, V^s_\infty+V^i_\infty, iV^s_\infty-iV^i_\infty]\in\CH$ have auxiliary components $\Bq_{\BE}$ and $\Bq_{\BJ}$ given by \eq{3.5fba}.
We require that the field components $\BE^s$ and $\BJ^s$ satisfy the constitutive law \eq{3.5}, while the remaining
auxiliary components satisfy the additional constitutive law
\beq \Bq_{\BJ}=\frac{i\Go}{\kappa_0}\Bq_{\BE}, \eeq{3.9a}
or equivalently, we have
\beq V^s_\infty+V^i_\infty=\Go(P^s_\infty-P^i_\infty)/\kappa_0,\quad iV^s_\infty-iV^i_\infty=i\Go(P^s_\infty+P^i_\infty)/\kappa_0.
\eeq{3.9ab}
The constitutive law \eq{3.5} allows us to relate the asymptotic terms of $\BJ^s(\Bx)$ and $\BE^s(\Bx)$ giving
\beq V^s_\infty=\Go P^s_\infty/\kappa_0,\quad V^i_\infty=\Go P^i_\infty/\kappa_0. \eeq{3.9ac}
In conjunction with \eq{3.9ab} this forces
\beq V^i_\infty(\Bn)= P^i_\infty(\Bn)=0, \eeq{3.9ad}
as desired. Thus we have replaced the Sommerfeld radiation condition with the constitutive law \eq{3.9a}. We arrive at a problem in the
extended abstract theory of composites: given $\Bs(\Bx)=(\BL_0-\BL(\Bx))\BE^a$, find fields $\BE$ and $\BJ$ in
the orthogonal spaces $\CE$ and $\CJ$ satisfying the constitutive law \eq{3.5} on $\CH^0$ and \eq{3.9a} on $\CA$.

\subsection{Expressing the acoustic scattered field in terms of integrals over the inclusion}
\labsect{farfieldac}

Let 
\beq {P^a}'(\Bx)=e^{i\Bk'_0\cdot\Bx} \quad {\rm and}~~ {\Bv^a}'(\Bx)=-i(\Go\Gr_0)^{-1}\Grad e^{i\Bk'_0\cdot\Bx}=\Bk'_0(\Go\Gr_0)^{-1} e^{i\Bk'_0\cdot\Bx} \eeq{4.17}
be another plane wave pressure and associated velocity field that solve the acoustic equations in the homogeneous medium
with density $\Gr_0$ and bulk modulus $\Gk_0$, i.e.
\beq
\underbrace{\begin{pmatrix}-i{\Bv^a}' \\ -i\Div{\Bv^a}' \end{pmatrix}}_{{\BJ^a}'}=
\underbrace{\begin{pmatrix}-(\Go\Gr_0)^{-1}\BI_d & 0 \\ 0 & \Go/\Gk_0\end{pmatrix}}_{\BL_0}
\underbrace{\begin{pmatrix}\Grad {P^a}' \\ {P^a}' \end{pmatrix}}_{{\BE^a}'}.
\eeq{3.12}
Using the key identity we have that 
\beqa I_1 & \equiv & (\BJ^s-\BL_0\BE^s,{\BE^a}')_r=(\BJ^s,{\BE^a}')_r-(\BE^s,\BL_0{\BE^a}')_r \nonum
& = & (\BJ^s,{\BE^a}')_r-(\BE^s,{\BJ^a}')_r
=\int_{\Md B_{r_0}}-i\overline{{P^a}'}\Bn\cdot\Bv^s-iP^s\Bn\cdot\overline{{\Bv^a}'}\,dS.
\eeqa{3.13}
Clearly the integrand on the left hand side vanishes outside $\GO$ and so the integral must be independent of the radius $r$ of the ball $B_{r_0}$
(so long as it contains the inclusion). So one can evaluate this integral by taking the limit $r_0\to\infty$. The
identity \eq{3.13} is the analog of the identity \eq{die8} that for the polarization problem expresses an integral over the inclusion in terms of the
far-field. 

Our goal is now to evaluate the integral on the right hand side of \eq{3.13} using the asymptotic formula,
\beq P^s(\Bx)=\frac{e^{ik_0|\Bx|}}{|\Bx|}P^s_\infty(\widehat{\Bx}),~~{\rm with}~ \widehat{\Bx}=\Bx/|\Bx|, \eeq{4.7}
for the scattered pressure field, and the associated asymptotic formula for the scattered velocity field 
$\Bv^s=-i(\Go\Gr_0)^{-1}\Grad P^s(\Bx)$. The calculation is the analog of the calculation \eq{die9}, that expresses
a far field integral in terms of the dipole moment.

Suppose we take a ball $ B$ of radius $r$. Then the outwards unit 
normal to the ball surface is $\Bn=\Bx/r$ and consequently $\Bn\cdot\Bx=r$.
Using the fact that $|\Bx|=\sqrt{\Bx\cdot\Bx}$ and $\widehat{\Bx}=\Bx/\sqrt{\Bx\cdot\Bx}$ this gives
\beq \Bn\cdot\Grad P^s(\Bx)=\frac{\Md P^s(\Bx)}{\Md r}
\approx \frac{ik_0e^{ik_0r}}{r}P^s_\infty(\widehat{\Bx})
-\frac{e^{ik_0r}}{r^2}P^s_\infty(\widehat{\Bx}).
\eeq{4.15}
Hence at large distances, keeping $\widehat{\Bx}$ fixed the dominant term in the above expression for $\Bn\cdot\Grad P^s(\Bx)$ is the first term.
So just keeping this, we obtain
\beq  \Bn\cdot\Bv^s=-i(\Go\Gr_0)^{-1}\Bn\cdot\Grad P^s(\Bx)\approx (\Go\Gr_0)^{-1}\frac{k_0e^{ik_0r}}{r}P^s_\infty(\widehat{\Bx}).
\eeq{4.16}
Recall the pressure field ${P^a}'(\Bx)$ and associated velocity field ${\Bv^a}'(\Bx)$ are given by \eq{4.17}.
So, we need to evaluate
\beq I_1=\int_{\Md B_{r_0}}-i\overline{{P^a}'}\Bn\cdot\Bv^s-iP^s\Bn\cdot\overline{{\Bv^a}'}\,dS
\approx -i(\Go\Gr_0)^{-1}\int_{\Md B_{r}}e^{-i\Bk'_0\cdot\Bx}e^{ik_0r}P^s_\infty(\widehat{\Bx})(k_0+\Bn\cdot\Bk'_0)/r\,dS.
\eeq{4.17a}
Without loss of generality let us suppose that the $x_1$ axis has been chosen in the direction of $\Bk'_0$, so $e^{i\Bk'_0\cdot\Bx}=e^{ik_0x_1}$
and $\Bn\cdot\Bk'_0=k_0n_1=k_0x_1/r$. Let us use cylindrical coordinates $(x_1,\Gvr,\Gt)$ where $\Gvr=\sqrt{x_2^2+x_3^2}$ and $\tan\Gt=x_3/x_2$,
so that $x_2=\Gvr\cos\Gt$ and  $x_3=\Gvr\sin\Gt$. We then introduce the ratio $t=x_1/r$ and express
\beq P^s_\infty(\widehat{\Bx})=P^s_\infty(\Gt,t). \eeq{4.18}
Thus in cylindrical coordinates the far-field expression for the scattered pressure field at $\Md B$ becomes
\beq P^s(\Bx)\approx\frac{e^{ik_0r}}{r}P^s_\infty(\Gt,x_1/r). \eeq{4.18a}
We choose as our variables of integration the parameters $t=x_1/r$ and $\Gt$. In terms of $t$ and $\Gt$, we have
\beqa &~& x_1=rt,\quad \Gvr=r\sqrt{1-t^2},\quad e^{-i\Bk'_0\cdot\Bx}=e^{-ik_0rt},\quad (k_0+\Bn\cdot\Bk'_0)/r=k_0(1+t)/r,
\nonum
&~& dS=\Gvr\,d\Gt\,dx_1/\sqrt{n_2^2+n_3^2}=r^2\,d\Gt\,dt,
\eeqa{4.19}
where $n_2$ and $n_3$ are the components of the vector $\Bn=\Bx/r$. The only term in the integration that involves $\Gt$ is $P^s_\infty(\Gt,h)$,
so integrating this over $\Gt$ defines
\beq p_{\infty}(t)\equiv\int_{0}^{2\pi}P^s_\infty(\Gt,t)\,d\Gt. \eeq{4.20}
We obtain
\beq I_1\approx -i(\Go\Gr_0)^{-1}I_2,\quad I_2=\int_{-1}^1rf(t)e^{irg(t)}\,dt, \eeq{4.21}
where
\beq g(t) = (1-t)k_0,\quad f(t) = k_0(1+t)p_{\infty}(t).
\eeq{4.22}
Asymptotic expressions in the limit $r\to\infty$ for integrals taking the form of $I_2$ in \eq{4.21} are available when $g(t)$ has a non-zero derivative
$g'(t)=k_0$ for $1\geq t \geq -1$, which is clearly the case, 
and one has \cite{Iserles:2005:EQH},
\beq \lim_{r\to\infty}I_2=\frac{e^{irg(1)}f(1)}{ig'(1)}-\frac{e^{irg(-1)}f(-1)}{ig'(-1)}. \eeq{4.24}
We have $g'(1)=g'(-1)=-k_0$, while \eq{4.22} and \eq{4.18a} imply
\beq g(1)=0,\quad g(-1)= 2k_0, \quad f(1)=2k_0p_{\infty}(1)=4k_0\pi P^s_\infty(\Bk'_0/k_0), \quad f(-1)=0, \eeq{4.25}
that when substituted in \eq{4.24} gives
\beq  \lim_{r\to\infty}I_2=4i\pi P^s_\infty(\Bk'_0/k_0), \eeq{4.26}
which is independent of $\Ga$ as expected.  Hence we obtain an exact expression for $I_1$:
\beq I_1=4k_0\pi P^s_\infty(\Bk'_0/k_0)/(\Go\Gr_0). \eeq{4.27}
Thus the scattered field $P^s_\infty(\Bn)$ can be determined from the integral $I_1$ over the inclusion,
given by \eq{3.13}.

There is also a connection, known as the acoustic ``optical theorem'', that links the forward scattering amplitude $P^s_\infty(\Bk_0/k_0)$
with the extinction $W$ which is the power taken out of the incident wave, thus corresponding to the sum of the power
absorbed and the power scattered:
\beq W  =  2k_0\pi\Imag[\overline{p^a} P^s_\infty(-\Bk_0/k_0)]/(\Go\Gr_0).
\eeq{6.14}
Proofs are given, for example, in \cite{Newton:1976:OTB, Dassios:2000:LFS}.

\section{Electromagnetic Scattering and Radiation}
\setcounter{equation}{0}
\labsect{esr}
The analysis here proceeds similarly to the acoustic case, the main difference being
the form of the fields $\BQ_{\BE}$ and  $\BQ_{\BJ}$. We start by considering plane wave
solutions $\Be^a$ and $\Bh^a$ for the electric and magnetic fields, respectively, in a
medium with electric permittivity $\Gve_0$ and magnetic permeability $\Gm_0$:
\beq \underbrace{\bpm i\Grad \times \Bh^a \\ -i\Bh^a \epm}_{\BJ_0}=
\underbrace{\begin{pmatrix} \Go \Gve_0 & 0 \\ 0 & -(\Go\Gm_0)^{-1}
  \end{pmatrix}}_{\BL_0}\underbrace{\bpm \Be^a \\ \Grad \times \Be^a \epm}_{\BE_0}.
\eeq{ems1}
With $\Be^a(\Bx)=\Be^a_0 e^{i\Bk_0\cdot\Bx}$ these have the solution:
\beq \BE^a=\bpm \Be^a \\ \Grad \times \Be^a \epm
=\begin{pmatrix} \Be_0^a \\ i\Bk_0\times\Be_0^a\end{pmatrix}e^{i\Bk_0\cdot\Bx}, \quad \quad
\BJ^a=\bpm i\Grad \times \Bh^a \\ -i\Bh^a \epm
=\bpm -\Bk_0 \times \Bh_0^a \\ -i\Bh_0^a \epm e^{i\Bk_0\cdot\Bx},
\eeq{ems2}
where $\Bh^a=\Bh_0^ae^{i\Bk_0\cdot\Bx}$ with
\beq \Bh^a_0=(\Go\Gm_0)^{-1}\Bk_0\times\Be^a_0,\quad
\Be^a_0=-(\Go\Gve_0)^{-1}\Bk_0\times\Bh^a_0. \eeq{ems3}
This implies $\Bk_0$ must be orthogonal to the real and imaginary parts
of $\Be^a_0$ with magnitude $k_0=|\Bk_0|$ given by
\beq k_0=\sqrt{\Go^2\Gve_0\Gm_0}. \eeq{ems4}

Suppose we are given fields 
\beq \BCP(\Bx)=\begin{pmatrix} \Bp_1(\Bx) \\ \Bp_2(\Bx) \end{pmatrix},
\quad \BCP'(\Bx)=\begin{pmatrix} \Bp'_1(\Bx) \\ \Bp'_2(\Bx) \end{pmatrix},
\eeq{ems6}
where $\Bp_1(\Bx)$, $\Bp_2(\Bx)$, $\Bp'_1(\Bx)$ and $\Bp'_2(\Bx)$
are $3$-dimensional vector fields. Their, possibly infinite, inner product is defined as
\beq (\BCP,\BCP')=\lim_{r_0\to\infty}\int_{t=0}^\infty w(t)(\BCP,\BCP')_{r_0t}\,\,dt, \eeq{ems7}
in which $w(t)$ is some smooth nonnegative weighting function, with say the properties that
\beq w(t)=0~~{\rm when}~~ t\leq 1/2~~{\rm or}~~t\geq 2,~~{\rm and}~~1=\int_{1/2}^2w(t)dt, \eeq{ems8}
and
\beq
(\BCP,\BCP')_r=\int_{B_r}\BCP(\Bx)\cdot\overline{\BCP'(\Bx)},\quad{\rm where}~
\BCP(\Bx)\cdot\overline{\BCP'(\Bx)}\equiv
\Bp_1(\Bx)\cdot\overline{\Bp_1'(\Bx)}+
\Bp_2(\Bx)\cdot\overline{\Bp_2'(\Bx)}, \eeq{ems9}
where $B_r$ is the ball of radius $r$,
and $\overline{\Ba}$ denotes the complex conjugate of $\Ba$ for any vector
quantity $\Ba$. We define $\CH^0$ as the space of 
fields $\BCP^0$ such that the norm $|h\BCP^0|=(h\BCP^0,h\BCP^0)^{1/2}$, with inner product given by \eq{3.5aa}, is finite for all scalar functions $h(\Bx)\in C_0^\infty(\R^d)$
(where $C_0^\infty(\R^d)$ is the set of all infinitely differentiable functions with compact support) and which additionally have the asymptotic behavior
\beq \BCP^0(\Bx)=\frac{e^{ik_0|\Bx|}}{|\Bx|}
\left\{\begin{pmatrix} \BR^s_\infty({\widehat{\Bx}})
    \\ \BS^s_\infty({\widehat{\Bx}}) \end{pmatrix}+
  \mathcal{O}\left(\frac{1}{|\Bx|}\right)\right\}
+\frac{e^{-ik_0|\Bx|}}{|\Bx|}\left\{\begin{pmatrix} \BR^i_\infty({\widehat{\Bx}}) \\ \BS^i_\infty({\widehat{\Bx}})
  \end{pmatrix}+\mathcal{O}\left(\frac{1}{|\Bx|}\right)\right\},
\eeq{ems10}
in which $\widehat{\Bx}=\Bx/|\Bx|$, and $\BR^s_\infty(\Bn)$,
$\BS^s_\infty(\Bn)$, $\BR^i_\infty(\Bn)$  and $\BS^i_\infty(\Bn)$
are complex vector functions defined on the unit sphere $|\Bn|=1$,
and tangential to it, i.e., satisfying
\beq \widehat{\Bx}\cdot\BR^s_\infty(\Bn)=0, \quad
     \widehat{\Bx}\cdot\BS^s_\infty(\Bn)=0,\quad
     \widehat{\Bx}\cdot\BR^i_\infty(\Bn)=0,\quad \text{ and} \quad
     \widehat{\Bx}\cdot\BS^i_\infty(\Bn)=0.
     \eeq{ems11}
Here the superscript $s$ is used because these field components will later be associated with the scattered field. The superscript $i$ is used because these field components will later be associated with 
incoming fields, though not the incoming fields associated with the incident fields $\Be^a$ and $\Bh^a$ as these will be treated separately. The subspace $\CH^0$ has been defined in this way to ensure that if $\BCP\in\CH^0$ then so is its complex conjugate  $\overline{\BCP}\in\CH^0$.
We define $\CH^s$ as the space of fields $\BCP^0\in\CH^0$ satisfying the 
condition that $\BR^i_\infty(\Bn)=\BS^i_\infty(\Bn)=0$ for all $\Bn$. Note that the norm $|\BCP^0|=(\BCP^0,\BCP^0)^{1/2}$ is not finite for fields in $\CH^0$ 
if $\BR^s_\infty(\Bn)$, $\BS^s_\infty({\widehat{\Bx}})$, $\BR^i_\infty(\Bn)$  or $\BS^i_\infty(\Bn)$ is nonzero. We define $\CH^0$ as the orthogonal complement of $\CV^0$ in the space $\CH^0$.

We are interested in solving
\beq \underbrace{\bpm i\Grad \times (\Bh^a+\Bh^s) \\ -i(\Bh^a+\Bh^s) \epm}_{\BJ^a+\BJ^s}
=
\underbrace{\begin{pmatrix} \Go \BGve & 0 \\ 0 & -(\Go\BGm)^{-1}\end{pmatrix}}_{\BL(\Bx)}
\underbrace{\bpm \Be^a+\Be^s \\ \Grad \times (\Be^a+\Be^s) \epm}_{\BE^a+\BE^s},
\eeq{ems12}
where $\Be^s(\Bx)$ and $\Bh^s(\Bx)$ are the scattered electric and magnetic fields and $\BJ^s,\BE^s\in\CH^s$. Here
the electric permittivity tensor $\BGve(\Bx)$ takes the value $\Gve_0\BI$ outside the inclusion,
and the magnetic permeability tensor $\BGm(\Bx)$ takes the value $\Gm_0\BI$ outside the inclusion. In a passive
possibly lossy system, at a non-zero, possibly complex, frequency with $\Imag\Go\geq 0$, then both $\Go \BGve$
and  $\Go \BGm$ (and hence $-(\Go\BGm)^{-1}$) have a positive semidefinite imaginary part, implying that
the imaginary part of $\BL(\Bx)$ is positive semidefinite. 

Now using the relation \eq{3.1}, that $\BJ^a=\BL_0\BE^a$, we rewrite \eq{ems12} as
\beq \BJ^s(\Bx)=\BL(\Bx)\BE^s(\Bx)-\Bs(\Bx),\quad \Bs(\Bx)=(\BL_0-\BL(\Bx))\BE^a. \eeq{ems13}
We define $\CE^0$ as the space of all fields $\BE^0$ in $\CH^0$ of the form
\beq \BE^0=\bpm \Be^0 \\ \Grad \times \Be^0 \epm, \eeq{ems14}
for some vector field $\Be^0(\Bx)$, and we define $\CJ^0$  as the space of all fields $\BJ^0$
in $\CH^0$ of the form
\beq\BJ^0= \bpm i\Grad \times \Bh^0 \\ -i\Bh^0 \epm,  \eeq{ems15}
for some vector field $\Bh^0(\Bx)$. The fields $\BJ^0$ and $\BE^0$, being in $\CH^0$, have the asymptotic 
forms
\beqa \BJ^0(\Bx) & = & \frac{e^{ik_0|\Bx|}}{|\Bx|}
\left\{\begin{pmatrix} -k_0\widehat{\Bx}\times\Bh^s_\infty({\widehat{\Bx}})
    \\ -i\Bh^s_\infty({\widehat{\Bx}}) \end{pmatrix}+
  \mathcal{O}\left(\frac{1}{|\Bx|}\right)\right\}
+\frac{e^{-ik_0|\Bx|}}{|\Bx|}\left\{\begin{pmatrix}
 k_0\widehat{\Bx}\times\Bh^i_\infty({\widehat{\Bx}})
    \\ -i\Bh^i_\infty({\widehat{\Bx}}) \end{pmatrix}
+\mathcal{O}\left(\frac{1}{|\Bx|}\right)\right\}, \nonum
\BE^0(\Bx)& = & \frac{e^{ik_0|\Bx|}}{|\Bx|}
\left\{\begin{pmatrix} \Be^s_\infty({\widehat{\Bx}})
    \\ ik_0\widehat{\Bx}\times\Be^s_\infty({\widehat{\Bx}}) \end{pmatrix}+
  \mathcal{O}\left(\frac{1}{|\Bx|}\right)\right\}
+\frac{e^{-ik_0|\Bx|}}{|\Bx|}
\left\{\begin{pmatrix} \Be^i_\infty({\widehat{\Bx}})
    \\ -ik_0\widehat{\Bx}\times\Be^i_\infty({\widehat{\Bx}}) \end{pmatrix}+
  \mathcal{O}\left(\frac{1}{|\Bx|}\right)\right\},
\eeqa{ems16}
implying, through \eq{3.5ba} and \eq{3.5bb}, that at large $|\Bx|$,
\beqa \Be^0(\Bx)& \approx & \frac{e^{ik_0|\Bx|}}{|\Bx|}\Be^s_\infty(\widehat{\Bx})+\frac{e^{-ik_0|\Bx|}}{|\Bx|}\Be^i_\infty(\widehat{\Bx}), \nonum
\Bh^0(\Bx)& \approx & \frac{e^{ik_0|\Bx|}}{|\Bx|}\Bh^s_\infty(\widehat{\Bx})+\frac{e^{-ik_0|\Bx|}}{|\Bx|}\Bh^i_\infty(\widehat{\Bx}),
\eeqa{ems17}
in which the asymptotic field components satisfy
\beqa \Be^s_\infty(\Bn)& = & -k_0\Bn\times\Bh^s_\infty(\Bn)/\Go\Gve_0, \quad \Bh^s_\infty(\Bn)=k_0\Bn\times\Be^s_\infty(\Bn)/\Go\Gm_0,\nonum
\Be^i_\infty(\Bn)& = &k_0\Bn\times\Bh^i_\infty(\Bn)/\Go\Gve_0, \quad \Bh^i_\infty(\Bn)=-k_0\Bn\times\Be^i_\infty(\Bn)/\Go\Gm_0,
\eeqa{ems17a}
where the relation $\Go^2\Gve_0\Gm_0=k_0^2$ ensures consistency of these relations.
The Silver-M{\"u}ller radiation condition that
\beqa {\widehat{\Bx}}\times\Bh^0& - &\sqrt{\frac{\Gve_0}{\Gm_0}}\Be^0\to 0\quad\text{as}\quad|\Bx|\to\infty, \nonum
(\text{or}\quad {\widehat{\Bx}}\times\Be^0& + &\sqrt{\frac{\Gm_0}{\Gve_0}}\Bh^0\to 0\quad\text{as}\quad|\Bx|\to\infty),
\eeqa{ems17b}
in fact implies that the $\Be^i_\infty(\widehat{\Bx})$ and $\Bh^i_\infty(\widehat{\Bx})$ associated with
the actual scattered electric and magnetic fields are zero, but we keep these terms 
as we want to impose a ``constitutive law at infinity'' that forces $\Be^i_\infty(\widehat{\Bx})$ and $\Bh^i_\infty(\widehat{\Bx})$ to be zero and thus
replaces the Silver-M{\"u}ller radiation condition.
Also we want to define the spaces $\CE^0$ and $\CJ^0$ so that if
$\BE^0$ and $\BJ^0$ are in  $\CE^0$ and $\CJ^0$, respectively, then so too are
$\Gl\BE^0$ and $\Gl\BJ^0$ for any complex constant $\Gl$.
We extend the definition of $\Be^s_\infty(\widehat{\Bx})$, and $\Bh^s_\infty(\widehat{\Bx})$
to all of $\R^3$ except the origin in the natural way by letting
\beq \Be^s_\infty(\Bx)=\Be^s_\infty(\Bx/|\Bx|),\quad \Bh^s_\infty(\Bx)=\Bh^s_\infty(\Bx/|\Bx|). \eeq{ems18}

Using integration by parts we have the key identity that
\beq (\BJ^0,\BE^0)_r=\int_{B_{r}} i\Bh^0\cdot(\Curl\overline{\Be^0})-i(\Curl\Bh^0)\cdot\overline{\Be^0}
\,d\Bx =\int_{B_{r}} -i\Div(\Bh^0\times\overline{\Be^0})\,d\Bx
=\int_{\Md B_{r}}-i\Bn\cdot(\Bh^0\times\overline{\Be^0})\,dS.
\eeq{ems19}
From \eq{3.5c} we see that when $|\Bx|$ is large,
\beq \Bh^0\times\overline{\Be^0}\approx
\frac{1}{|\Bx|^2}\left(\Bh^s_\infty\times\overline{\Be^s_\infty}+\Bh^i_\infty\times\overline{\Be^i_\infty}\right)
+\frac{e^{2ik_0r}}{|\Bx|^2}\Bh^s_\infty\times\overline{\Be^i_\infty}
+\frac{e^{-2ik_0r}}{|\Bx|^2}\Bh^i_\infty\times\overline{\Be^s_\infty}.
\eeq{ems20}
The last two cross terms that involve $e^{2ik_0r}$ and $e^{-2ik_0r}$ obviously oscillate very rapidly with $r$ and will average to zero
in the integral \eq{ems7} involving the smooth weight function $w(t)$. Thus we get
\beqa (\BJ^0,\BE^0)& = & \lim_{r_0\to\infty}\int_{t=0}^\infty dt \frac{-iw(t)}{r_0^2}\int_{\Md B_{r_0t}}
\Bn\cdot\left[\Bh^s_\infty(\widehat{\Bx})\times\overline{\Be^s_\infty(\widehat{\Bx})}+\Bh^i_\infty(\widehat{\Bx})\times\overline{\Be^i_\infty(\widehat{\Bx})}\right]\,dS \nonum
& = &-i\int_{\Md B_{1}}
\Bn\cdot\left[\Bh^s_\infty(\Bn)\times\overline{\Be^s_\infty(\Bn)}+\Bh^i_\infty\times\overline{\Be^i_\infty(\Bn)}\right]\,dS,
\eeqa{ems21}
This lack of orthogonality of the subspaces $\CE^0$ and $\CJ^0$ can be remedied by introducing an auxiliary space $\CA$ of 
four-component fields $\BQ(\Bn)=[\Bq_1(\Bn),\Bq_2(\Bn),\Bq_3(\Bn),\Bq_4(\Bn)]$, in which the $\Bq_i(\Bn)$ are vector fields
defined, and square integrable, on the unit sphere $|\Bn|=1$. Additionally, we require that  $\Bn\cdot\Bq_i(\Bn)=0$ for $i=1,2,3,4$,
and for all $\Bn$ (so that the fields are tangential to the sphere).
We then consider the Hilbert space $\CH$ composed of fields $[\BCP,\BQ]$, where $\BCP\in\CH^0$
and $\BQ=[\Bq_1,\Bq_2,\Bq_3,\Bq_4]\in\CA$.
In general, the field components $\Bq_i(\Bn), i=1,2,3,4$ need not be related to the
functions $\BR^s_\infty(\Bn)$, $\BS^s_\infty({\widehat{\Bx}})$, $\BR^i_\infty(\Bn)$  and $\BS^i_\infty(\Bn)$
appearing in the asymptotic expansion \eq{ems10}. The inner product
between two fields $\BCQ=[\BCP,\BQ]$ and $\BCQ'=[\BCP',\BQ']$ in $\CH$ is defined as
\beq \lang\BCQ,\BCQ'\rang =(\BCP,\BCP')+\frac{1}{2}\sum_{i=1}^4\int_{|\Bn|=1}\Bq_i(\Bn)\cdot\overline{\Bq'_i(\Bn)}\,dS. \eeq{ems21a}
We define $\CE$ to consist of fields $\BE=(\BE^0,\BQ_{\BE})$, and  $\CJ$ to consist of fields $\BJ=(\BJ^0,\BQ_{\BJ})$, where
$\BE^0\in\CE^0$ and $\BJ^0\in\CJ^0$, while $\BQ_{\BE}$ and $\BQ_{\BJ}$ take the form
\beq \BQ_{\BE} = \bpm -i\Be^s_\infty+i\Be^i_\infty\\ \Be^s_\infty+\Be^i_\infty \\
      \Bn\times\Be^s_\infty-\Bn\times\Be^i_\infty \\i\Bn\times\Be^s_\infty+i\Bn\times\Be^i_\infty\epm\in\CH,\quad
\BQ_{\BJ} = \bpm \Bn\times\Bh^s_\infty-\Bn\times\Bh^i_\infty\\ i\Bn\times\Bh^s_\infty+i\Bn\times\Bh^i_\infty\\
-i\Bh^s_\infty+i\Bh^i_\infty\\ \Bh^s_\infty+\Bh^i_\infty\epm\in\CH,
\eeq{ems22}
where $\Be^s_\infty(\Bn)$, $\Be^i_\infty(\Bn)$, $\Bh^s_\infty(\Bn)$ and $\Bh^i_\infty(\Bn)$ are those functions that enter the
asymptotic forms \eq{ems16}.
The auxiliary fields  $\BQ_{\BE}$ and $\BQ_{\BJ}$ defined in this way ensure that if $\BJ\in\CJ$ and $\BE\in\CE$
then also $\overline{\BJ}\in\CJ$ and $\overline{\BE}\in\CE$. When $\BE^0=\BE^s$ and $\BJ^0=\BJ^s$, where $\BE^s$ and $\BJ^s$
satisfy the constitutive relation \eq{ems13},
we require that their auxiliary field components satisfy the additional constitutive relation
\beq \BQ_{\BJ}=\frac{ik_0}{\Go\Gve_0}\BQ_{\BE}, \quad\text{or equivalently} \quad \BQ_{\BJ}=\frac{i\Go\Gm_0}{k_0}\BQ_{\BE},
\eeq{ems23}
that, together with the relations \eq{ems17a}, forces $\Be^i_\infty(\Bn)=\Bh^i_\infty(\Bn)=0$ and thus replaces the
Silver-M{\"u}ller radiation condition.

Now the inner product of $\BJ$ and $\BE$ is 
\beqa \lang \BJ,\BE\rang& = & (\BJ^0,\BE^0)+i\int_{\Md B_{1}}(\Bn\times\Bh^s_\infty)\cdot\overline{\Be}^s_\infty
-\Bh^s_\infty\cdot(\Bn\times\overline{\Be}^s_\infty)
+(\Bn\times\Bh^i_\infty)\cdot\overline{\Be}^i_\infty
-\Bh^i_\infty\cdot(\Bn\times\overline{\Be}^i_\infty)\,dS \nonum
& = & (\BJ^0,\BE^0)+i\int_{\Md B_{1}}\Bn\cdot(\Bh^s_\infty\times\overline{\Be}^s_\infty)+\Bn\cdot(\Bh^i_\infty\times\overline{\Be}^i_\infty)\,dS=0,
\eeqa{ems24}
which, with \eq{ems21}, implies the orthogonality of the spaces $\CE$ and $\CJ$. We again have arrived at a problem in the
extended abstract theory of composites: given $\Bs(\Bx)=(\BL_0-\BL(\Bx))\BE^a$, find fields $\BE$ and $\BJ$ in
the orthogonal spaces $\CE$ and $\CJ$ satisfying the constitutive law \eq{ems13} on $\CH^0$ and \eq{ems23} on $\CA$.
\subsection{Expressing the electromagnetic
  scattered field in terms of integrals over the inclusion}
\labsect{farfieldel}
Let us consider another plane wave solution for the electric and magnetic fields in a
medium with electric permittivity $\Gve_0$ and magnetic permeability $\Gm_0$,
\beq {\Be^a}'(\Bx)={\Be^a_0}' e^{i\Bk_0'\cdot\Bx},\quad {\Bh^a}'={\Bh_0^a}'e^{i\Bk_0'\cdot\Bx},\quad 
{\Bh^a_0}'=(\Go\Gm_0)^{-1}\Bk_0'\times{\Be^a_0}',\quad
{\Be^a_0}'=-(\Go\Gve_0)^{-1}\Bk_0'\times{\Bh^a_0}'. \eeq{si1}
These solve
\beq \underbrace{\bpm i\Grad \times \Bh^a \\ -i\Bh^a \epm}_{\BJ_0}=
\underbrace{\begin{pmatrix} \Go \Gve_0 & 0 \\ 0 & -(\Go\Gm_0)^{-1}
  \end{pmatrix}}_{\BL_0}\underbrace{\bpm \Be^a \\ \Grad \times \Be^a \epm}_{\BE_0}.
\eeq{si2}
Using the key identity we have that 
\beqa I_1 & \equiv & (\BJ^s-\BL_0\BE^s,{\BE^a}')_r=(\BJ^s,{\BE^a}')_r-(\BE^s,\BL_0{\BE^a}')_r \nonum
& = & (\BJ^s,{\BE^a}')_r-(\BE^s,{\BJ^a}')_r
=\int_{\Md B_{r}}-i\Bn\cdot(\Bh^s\times\overline{{\Be^a}'})-i\Bn\cdot(\overline{{\Bh^a}'}\times\Be^s)
\,dS.
\eeqa{si3}
At large $|\Bx|$ we have
\beqa
\Be^s(\Bx) & \approx &\frac{e^{ik_0|\Bx|}}{|\Bx|}\Be^s_\infty(\widehat{\Bx}),\quad \Bh^s(\Bx) \approx  \frac{e^{ik_0|\Bx|}}{|\Bx|}\Bh^s_\infty(\widehat{\Bx}),\nonum
 \Be^s_\infty(\Bn)& = & -k_0\Bn\times\Bh^s_\infty(\Bn)/\Go\Gve_0, \quad \Bh^s_\infty(\Bn)=k_0\Bn\times\Be^s_\infty(\Bn)/\Go\Gm_0,
 \eeqa{si4}
 and so $I_1$ is given by
 \beq I_1\approx -i\int_{\Md B_{r}}w(\Bn)e^{i(k_0r-\Bk_0'\cdot\Bx)}/r
\,dS, \quad\text{where}\quad w(\Bn)=\Bn\cdot [\Bh^s_\infty(\Bn)\times\overline{{\Be^a_0}'(\Bn)}+\overline{{\Bh^a_0}'(\Bn)}\times\Be^s_\infty(\Bn)].
\eeq{si5}
In the special case when $\Bn=\Bk_0'/k_0$, where $\Be^s_\infty$ and $\overline{{\Be^a_0}'}$ are both
perpendicular to $\Bn$, we get
\beq  w(\Bn)=k_0(\Go\Gm_0)^{-1}\Bn\cdot [(\Bn\times\Be^s_\infty)\times\overline{{\Be^a_0}'}
+(\Bn\times\overline{{\Be^a_0}'})\times\Be^s_\infty]=-2k_0(\Go\Gm_0)^{-1}(\Be^s_\infty\cdot\overline{{\Be^a_0}'}),
\eeq{si5a}
and, similarly, $w(\Bn)=0$ when $\Bn=-\Bk_0'/k_0$

With cylindrical coordinates $(x_1, \Gvr, \Gt)$ where $x_1$ is in the direction of $\Bk_0'$, $\Bn=\widehat{\Bx}$ just depends on $\Gt$
and $t=x_1/r$. Defining
\beq f(t)=\int_0^{2\pi}w(\Bn),d\Gt,\quad g(t) = (1-t)k_0,
\eeq{si6}
we obtain
\beq I_1\approx -i\int_{-1}^1rf(t)e^{irg(t)}\,dt. \eeq{si7}
Using the asymptotic expression \eq{4.24} for the integral, and noting that
\beq g'(1)=g'(-1)=-k_0,\quad g(1)=0,\quad g(-1)= 2k_0, \quad
f(1)=-4\pi k_0(\Go\Gm_0)^{-1}(\Be^s_\infty\cdot\overline{{\Be^a_0}'}),\quad f(-1)=0,
\eeq{si8}
where the formula for $f(1)$ follows from \eq{si5a}, gives 
\beq I_1=-4\pi(\Go\Gm_0)^{-1}(\Be^s_\infty(\Bk_0'/k_0)\cdot\overline{{\Be^a_0}'})
\eeq{si9}
As $\overline{{\Be^a_0}'}$ can be any vector perpendicular to $\Bk_0'$, and since
$\Be^s_\infty(\Bk_0'/k_0)$ is also perpendicular to $\Bk_0'$, we can recover the scattering amplitudes
$\Be^s_\infty(\Bn)$ from such integrals for all vectors $\Bn$ with $|\Bn|=1$.

\section*{Acknowledgments}
GWM thanks the National Science Foundation for support through grants DMS-1814854 and DMS-2107926 . The work is partly based on the
books \cite{Milton:2002:TOC, Milton:2016:ETC} and again I thank those (cited in the acknowledgments of Part I) who
helped stimulate that work and who provided feedback on the drafts of those books. The work is additionally based
on the paper \cite{Milton:2017:BCP}.
Nelson Beebe is thanked for all the work he did on preparing the books for publication and for updating the associated bibtex entries.
Yury Grabovsky is thanked for helpful comments.
\ifx \bblindex \undefined \def \bblindex #1{} \fi\ifx \bbljournal \undefined
  \def \bbljournal #1{{\em #1}\index{#1@{\em #1}}} \fi\ifx \bblnumber
  \undefined \def \bblnumber #1{{\bf #1}} \fi\ifx \bblvolume \undefined \def
  \bblvolume #1{{\bf #1}} \fi\ifx \noopsort \undefined \def \noopsort #1{}
  \fi\ifx \bblindex \undefined \def \bblindex #1{} \fi\ifx \bbljournal
  \undefined \def \bbljournal #1{{\em #1}\index{#1@{\em #1}}} \fi\ifx
  \bblnumber \undefined \def \bblnumber #1{{\bf #1}} \fi\ifx \bblvolume
  \undefined \def \bblvolume #1{{\bf #1}} \fi\ifx \noopsort \undefined \def
  \noopsort #1{} \fi

\end{document}